\documentclass[a4paper]{article}

\usepackage{INTERSPEECH2020}
\usepackage{multirow,comment}
\title{Speaker Re-identification with Speaker Dependent Speech Enhancement}

\name{Yanpei Shi, Qiang Huang, Thomas Hain}

\address{ Speech and Hearing Research Group\\
          Department of Computer Science, University of Sheffield}
\email{\{YShi30, qiang.huang, t.hain\}@sheffield.ac.uk}

\begin{document}

\maketitle
\begin{abstract}
While the use of deep neural networks has significantly boosted speaker recognition performance,
it is still challenging to separate speakers in poor acoustic environments. 
Here speech enhancement methods have traditionally allowed improved performance. 
The recent works have shown that adapting speech enhancement can lead to further gains. 
This paper introduces a novel approach that cascades speech enhancement and 
speaker recognition. In the first step, a speaker embedding vector is generated
, which is used in the second step to enhance the speech quality and re-identify the speakers.
Models are trained in an integrated framework with joint optimisation.
The proposed approach is evaluated using the Voxceleb1 dataset, which aims
to assess speaker recognition in real world situations. In 
addition three types of noise at different signal-noise-ratios were added
for this work. 
The obtained results show that the proposed approach using speaker
dependent speech enhancement can yield better speaker recognition and speech enhancement performances
than two baselines in various noise conditions.
\end{abstract}
\noindent\textbf{Index Terms}: Speech Enhancement, Speaker Identification, Speaker Verification, Noise Robustness.

\vspace*{-2mm}
\section{Introduction}\label{introduction}
\vspace*{-2mm}

The aim of speaker recognition is to recognize speaker identities from their voice characteristics \cite{poddar2017speaker}. 
In recent years, the use of deep learning technologies \cite{variani2014deep, snyder2018x, rahman2018attention,an2019deep} has
significantly improved speaker recognition performance.
However, speaker recognition in poor noise conditions is still
a challenging task as some important acoustic information related to the speaker is often
interfered.
To tackle speech signals corrupted by noise, some methods have been developed.
Previous studies \cite{leglaive2019speech,sadeghi2019audio,zhao2014robust} tended to recover original signals by removing
noise. Other methods \cite{jang2017enhanced,farahani2006robust,ming2007robust} focused on feature extraction from un-corrupted speech signals,
and further methods \cite{nahma2019adaptive,yao2016priori} tried to estimated speech quality by computing signal-to-noise ratios (SNRs).



In many previous studies, speech enhancement is often processed individually \cite{ouyang2019fully,pascual2017segan,muhammed2019non,yao2019coarse}. 
However,
the learned features or enhanced speech signals might not be able to match well to
the information required by speaker recognition and verification.
It is highly desirable that both the speech enhancement and the speaker processing models can work together and
can be optimized jointly. 
In \cite{shon2019voiceid}, Shon et al. tried to integrate speech enhancement module and speaker processing
module into one framework.
In this method, a speech enhancement module filters out the noise
by generating a ratio mask, and then multiplying it by the original spectrogram.
However, in \cite{shon2019voiceid}, the speaker verification module was pertrained and fixed when training  
the speech enhancement.
The two modules are not optimized jointly.

\begin{figure*}[h]
	\centering
	\includegraphics[height=4.5cm,width=15cm]{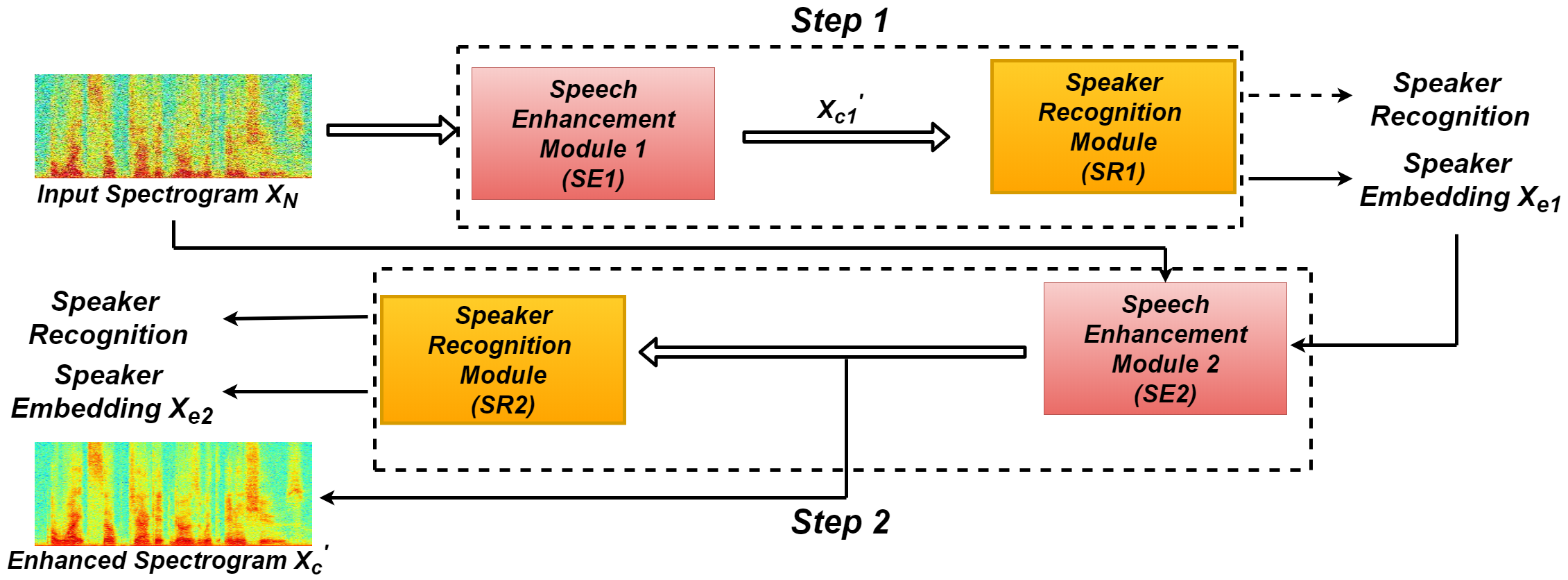}
	\caption{Architecture of the proposed approach consisting of two steps (Step1 and Step2), 
	         each of which contains two modules: a speech enhancement (\textit{SE}) module and a speaker recognition (\textit{SR}) module. 
	         The input is a spectrogram corrupted by noise. A speaker identity and an enhanced spectrogram are the output.}
	\label{overall}
\end{figure*}

To improve speaker identification and verification performance,  
our proposed approach proposes a
speech enhancement module cascaded to a speaker
recognition module in order to reduce the impact caused by noise interference.
The two modules
are optimized jointly
by computing the enhancement loss and identification loss simultaneously.

To our knowledge, although speaker information has been widely used for
acoustic model adaptation in speech recognition \cite{meng2019speaker,zhao2018domain}, it is still under-developed
in speech enhancement.
To further improve the robustness of speaker recognition against noise,
two steps are taken in this work. The first step is to
learn a speaker embedding vector, which will be then
used as a prior knowledge to enhance speech quality in the second step.
The details of the proposed approach will be described in the following sections.

The rest of the paper is organized as follow: 
Section \ref{model architecture} presents the model architecture of the proposed approach 
and how it is implemented in order to identify a speaker and enhance the speech quality simultaneously.
The used data set and experiment set-up are introduced in Section \ref{Experiments}. 
The obtained results and related analysis are given in Section \ref{results}, 
and finally conclusions are drawn in Section \ref{conclusion}.

\vspace*{-3mm}
\section{Speaker Re-Identification}\label{model architecture}
\vspace*{-2mm}
%

\subsection{Model Structure}
\vspace*{-2mm}
Figure \ref{overall} shows the architecture of the proposed approach, consisting of two steps (Step1 and Step2). 
Each step contains two modules, a speech enhancement (\textit{SE}) module and 
a speaker recognition (\textit{SR}) module. Given an input spectrogram $\boldsymbol {X_{N}}$, the goal for Step1 is to generate a speaker embedding ($X_{e1}$) using the speech enhancement module (\textit{SE}1) and speaker recognition module (\textit{SR}1). 
In Step2, the speaker embedding $X_{e1}$ is used as the prior knowledge to improve the speaker recognition and speech enhancement performances. 
The architecture of the speech enhancement module (\textit{SE}2) and speaker recognition module (\textit{SR}2) have similar architecture to
the \textit{SE}1 and \textit{SR}1 modules in Step1. The only difference is \textit{SE}2 takes $X_{e1}$ into account.

\vspace*{-3mm}
\subsection{Module of Speech Enhancement and Speaker Recognition}
\vspace*{-2mm}

\begin{figure}[tbh]
	\centering
	\includegraphics[height=2.5cm,width=8cm]{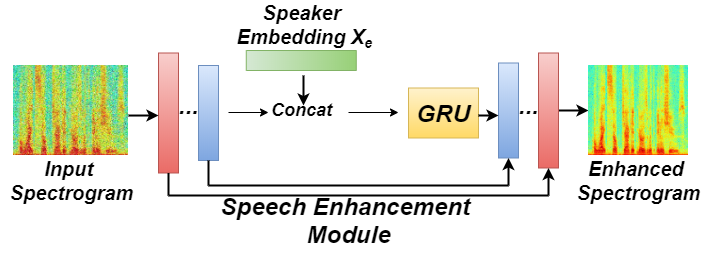}	
	
	\caption{Structure of the speech enhancement (\textit{SE}) model is built on a residual/skip auto-encoder network and used in both Step1 and Step2.
	  The speaker embedding is used only in \textit{SE}2.}
	\label{fig:se}
\end{figure}

Figure \ref{fig:se} shows the structure of
the speech enhancement (\textit{SE}) module.
It is based on the structure of a residual auto-encoder \cite{pandey2019new,pascual2017segan,leglaive2019speech}, where several convolutional layers are stacked. 
It can be viewed as one stack of several singlelayer auto-encoders. The residual connection could improve the quality of the reconstructed spectrogram and avoid the
vanishing gradient problem \cite{pandey2019new}. The use of a bi-directional
GRU layer inserted between the encoder
and decoder is to improve performance of speech enhancement
\cite{liu2018denoising}, as it takes context information into account. The speaker
embedding is only used in SE2.

\begin{figure}[tbh]
	\centering
	\includegraphics[height=2.5cm,width=8cm]{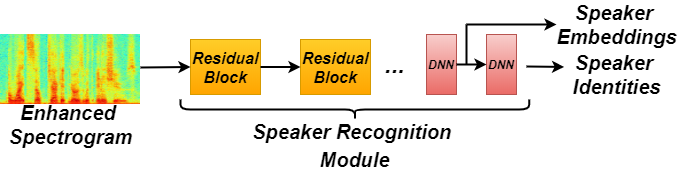}	
	
	\caption{Structure of speaker recognition (\textit{SR}) module is built on a Resnet-20 network. 
	SR1 aims to generate a speaker embedding, and \textit{SR}2 is in charge of recognizing speaker ID.}
	\label{fig:sr}
\end{figure}

Figure \ref{fig:sr} shows the structure of the used \textit{SR} module, which is built on a Resnet-20 \cite{hajibabaei2018unified} structure.
The following two fully-connected (FC)
layers are used for speaker classification, and the output
of the second to last FC layer is defined as a speaker embedding.

\vspace*{-2mm}
\subsection{Speaker Embeddings (Step1)}
\vspace*{-2mm}
As shown in Figure \ref{overall}, \textit{SE}1 and \textit{SR}1 are cascaded.
The noise corrupted spectrogram $\boldsymbol {X_{N}}$ is denoised using \textit{SE}1, by which a speaker embedding is
then yielded by the first fully connected layer of \textit{SR}1.

For the \textit{SE}1 module, mean absolute error (MAE) \cite{willmott2005advantages} is used to measure 
the difference between an input spectrogram $\boldsymbol {X_{C}}$ and an enhanced spectrogram $\boldsymbol {X_{C1}^{'}}$, 
as it is more efficient compared to mean squared error (MSE) \cite{pandey2019new}.
The loss function of this module is defined as:

\begin{equation}\label{se_train}
\footnotesize
\boldsymbol {\mathcal{L}_{SE}} = \frac{1}{TF}||\boldsymbol {X_{C}} - \boldsymbol {X_{C1}^{'}}||_{1} =  \frac{1}{TF} \sum_{i=0}^{T}\sum_{j=0}^{F}|x_{ij} - x^{'}_{ij}|
\end{equation}
where $x_{ij}$ and $x^{'}_{ij}$ denotes each element in the clean and denoised spectrogram. 
$T$ and $F$ denote the dimension on time and frequency axes, respectively.

For SR1 module, the classifier is trained in terms of the difference between predictions $\hat{y}$
and corresponding targets $y$ and uses the categorical cross entropy as its loss function:  

\begin{equation}\label{sid_train}
\footnotesize
\boldsymbol {\mathcal{L}_{SR}} = - \sum_{i=0}^{N}\sum_{j=0}^{M}y_{ij}* \log \hat y_{ij}
\end{equation}

where $N$ denotes the number of samples and $M$ denotes the number of classes (speakers). 

The \textit{SE}1 and the \textit{SR}1 modules are firstly trained independently using the loss function introduced above and then fine-tuned together using Eq \ref{fig:sr}. 


\vspace*{-2.5mm}
\subsection{Speaker-Dependent Speech Enhancement (Step2)}
\vspace*{-2mm}
In Step2, both the speech enhancement and speaker recognition modules
use similar structures to the modules in Step1. However, unlike \textit{SE}1, the \textit{SE}2 module concatenates the speaker embedding vector $\boldsymbol {X_{e1}}$,
with its own bottleneck vector and enhances the quality of $\boldsymbol {X_{N}}$.

In this work, the optimization of Step1 and Step2
are independent to each other. The parameters of \textit{SE}1 and \textit{SR}1
used in Step1 are fixed when training \textit{SE}2. \textit{SR}2 shares weights with \textit{SR}1. 
Unlike Step1, a joint optimization is implemented on 
the two modules in Step2 by using
Eq \ref{se_train} and \ref{sid_train} simultaneously:
\begin{equation}\label{se_sr_train}
\footnotesize
   \boldsymbol {\mathcal{L}} = \boldsymbol {\mathcal{L}_{SE}}  + \boldsymbol {\mathcal{L}_{SR}} 
\end{equation}

\vspace*{-3mm}
\section{Experiments}\label{Experiments}
\vspace*{-3mm}
\subsection{Data}\label{Data}
\vspace*{-2mm}
All experiments were run on the Voxceleb1 dataset \cite{nagrani2017voxceleb},
which is a widely used large dataset for speaker identification and verification. 
The Voxceleb1 dataset contains 1251 speakers and more than 150 thousand
``wild'' utterances,
extracted from YouTube videos.

In all experiments, spectrograms are used as the input acoustic features. 
Input speech streams were firstly segmented using a 25ms sliding window with a 10ms step. 
A 512-point Fast Fourier Transform (FFT) was then conducted on each segment which yielded a 257-D vector (a DC component is concatenated). 
The length of spectrogram covers 300 frame, about 3 seconds. 
No normalization techniques were used to pre-process the spectrograms.

To evaluate the robustness of the proposed model, additional noises from the MUSAN dataset were added. 
The MUSAN dataset contains three categories of noises: general noise, music and babble \cite{snyder2015musan}. 
The general noise type contains 6 hours of audio, including dialtones and fax machine noises etc. 
The music data contains 42 hours of music recordings from different categories, and
the babble data contains 60 hours of speech, including read speech from public domain, hearings, committees and debates etc.

\vspace*{-3mm}
\subsection{Experiment Setup}\label{experiments setup}
\vspace*{-2mm}
To evaluate the effectiveness of the proposed approach,
two tasks, speaker identification and verification, were designed and tested on the Voxceleb1 dataset
using the official train/test split \cite{nagrani2017voxceleb}. 

For speaker identification, there are 1251 speakers in both training and test set \cite{nagrani2017voxceleb}. 
Each utterance is randomly mixed with a type of noise at one of five SNR levels (from 0 to 20 dB). 
To evaluate the recognition performance, The Top-1 and Top-5 accuracies were computed \cite{hajibabaei2018unified}.

For speaker verification, the same data configuration as the speaker identification task was set. 
A cosine score between two vectors was computed and used to measure the similarity \cite{shon2019voiceid}. 
Equal Error Rate (EER) \cite{cheng2004method} and Detection Cost Function (DCF) \cite{van2007introduction} 
were used as evaluation metrics.
DCF represents the average of two minimum DCF score (DCF0.01 and DCF0.001) \cite{van2007introduction,xie2019utterance}.


To evaluate our proposed approaches, two baselines and two proposed approaches were tested. 
\begin{description}
\item[\textbf{SID}]: represents the baseline method using only a speaker recognition module (\textit{SR}1) without any pre-processing and post-processing. 
\item[\textbf{VoiceID-Loss}]\cite{shon2019voiceid}: represents a baseline from \cite{shon2019voiceid}, where the
speech enhancement and speaker recognition modules are cascaded, but without a joint training and the use of speaker embeddings.
\item[\textbf{SESR-Step1}]: represents the proposed model where the \textit{SE}1 and \textit{SR}1 modules are jointly trained, but 
speaker embedding vectors are not being used.
\item[\textbf{SESR2-Step2}]: represents the model where the \textit{SE}2 and \textit{SR}2 modules are jointly trained 
with the learned speaker embedding vector being used in \textit{SE}2. The loss function, defined by Eq \ref{se_sr_train}), is employed
for model optimization.
\end{description}

To verify the effectiveness of the proposed approach in speech enhancement,
two metrics,
Perceptual Evaluation of Speech Quality (PESQ) \cite{rix2001perceptual} and
Short-Time Objective Intelligibility (STOI) \cite{taal2010short}, are used
to assess the enhanced speech quality.

  
\vspace*{-2mm}
\subsection{Network Structure}
\vspace*{-3mm}

\begin{table}[h]
	\renewcommand\arraystretch{1.0}
	\centering  
	\footnotesize
	\begin{tabular}{c|c|c|c}
		\hline
		Operation&Structure& Input (T, F, C) & Output (T, F, C) \\
		\hline
		\multirow{5}{*}{Encoder}&
		16/(1,2) & (300,257,1) & (300,129,16)\\
		&32/(2,2) &(300,129,16)& (150,65,32)\\
		&64/(2,2) & (150,65,32)& (75,33,64)\\
		&128/(2,2) & (75,33,64)& (38,17,128)\\
		&256/(2,4) & (38,17,128)& (19,5,256)\\
		\hline
		Reshape &- &(19,5,256)& (19,1280)\\
		Concatenation &- &(19,1280)& (19,1536)\\
		DNN &512 &(19,1536)& (19,512)\\
		Bi-GRU&640&(19,512)&(19,1280)\\
		Reshape &- &(19,1280)& (19,5,256)\\
		\hline

	\end{tabular}
	\caption{The encoder architecture of the proposed speech enhancement approach, where $T$, $F$, $C$ represents the time, frequency and feature dimensions. The number of features and strides on each dimension are shown as Feature/Strides}\label{model_summary}
	\label{model_sum}
\end{table}

\begin{figure}[h]
	\begin{minipage}{0.48\linewidth}
		\centering
		\includegraphics[width=4.0cm,height=2.6cm]{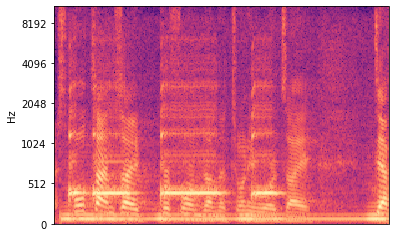}
		\centerline{  (a) Noise speech}
	\end{minipage}
	\begin{minipage}{0.48\linewidth}
		\centering
		\includegraphics[width=4.0cm,height=2.6cm]{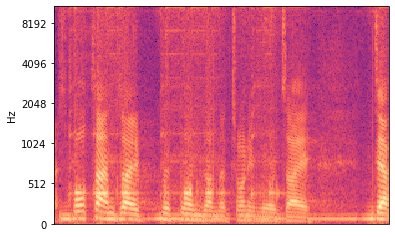}
		\centerline{  (b) Enhanced speech using SE1}
	\end{minipage}

	\begin{minipage}{0.48\linewidth}
		\centering
		\includegraphics[width=4.0cm,height=2.6cm]{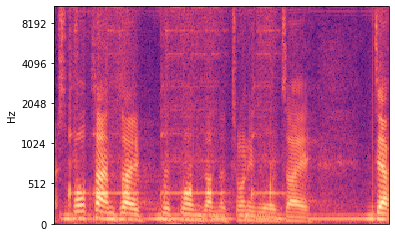}
		\centerline{  (c) Enhanced speech using SE2}
	\end{minipage}
	\begin{minipage}{0.48\linewidth}
		\centering
		\includegraphics[width=4.0cm,height=2.6cm]{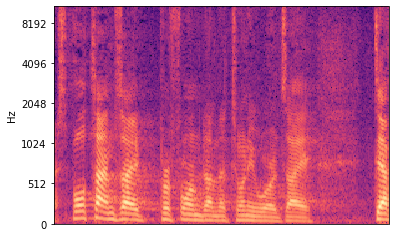}
		\centerline{  (d) Original speech}
	\end{minipage}
\caption{Comparison of spectrograms (a) speech corrupted by noise; (b) enhanced spectrogram obtained using the SE1 module; (c) enhanced spectrogram obtained using the SE2 module; (d) original speech.}\label{fig:enhanced-spec}
\end{figure} 
 
Table \ref{model_sum} shows the encoder architecture of the skip/residual auto-encoder employed by
the speech enhancement module used in Step1 and Step2. 
Its decoder structure mirrors the encoder.  
For the speaker recognition module, the structure of Resnet-20 is used and the details can be found in \cite{he2016deep}.

For model optimization, The Adam optimizer \cite{kingmaadam} is used with the initial learning rate
being set to 1e-3 and the decay rate being set to 0.9 for each epoch.

\begin{table*}[h]
	\renewcommand{\multirowsetup}{\centering}  
	\renewcommand\arraystretch{1.0}
	\setlength{\tabcolsep}{2mm}
	\centering  
	\footnotesize
	\begin{tabular}{c|c|c|c|c|c|c|c|c|c}
		\hline
		\multirow{2}{*}{\textbf{Noise Type}}& \multirow{2}{*}{\textbf{SNR}} & \multicolumn{2}{c|}{\textbf{SID}}&\multicolumn{2}{c|}{\textbf{VoidID-Loss\cite{shon2019voiceid}}} & 
		\multicolumn{2}{c|}{\textbf{SESR-Step1}}&
		\multicolumn{2}{c}{\textbf{SESR-Step2}}\\
		\cline{3-10}
		& & Top1 (\%)&Top5 (\%)&Top1 (\%)&Top5 (\%)& Top1 (\%)& Top5 (\%)&Top1 (\%)& Top5 (\%)\\

		\hline
		\multirow{5}{*}{\textbf{Noise}}&
		0 &74.1 &86.9 &75.6 & 88.0 & 76.4 & 88.9  & 77.4& 89.4\\
		&5& 79.2&90.0 &80.4 & 90.8& 81.8& 91.2  & 83.6&91.6\\
		&10&83.2 &93.2 &84.7 &94.3 & 85.4 & 94.7 & 87.1 &95.7\\
		&15&84.9 &94.6 & 85.6& 95.1& 86.3 &95.8  & 88.7 &95.9\\
		&20&87.9 &95.4 &88.7 & 96.0& 89.5 & 96.4 & 90.3 &96.8\\
		\hline
		
		\multirow{5}{*}{\textbf{Music}}&
		0 & 65.8&82.0 &67.1 &83.3 & 69.2 & 85.2 &70.4 &85.8\\
		&5&76.9 &89.1 &78.2 & 89.9&  80.1 & 90.6 &81.3 &90.6\\
		&10&83.8 &93.5 & 84.6 & 94.2& 85.9 &  95.1&85.9 &95.5\\
		&15&86.1 & 93.9&87.3 & 95.0& 88.4  & 95.7 & 89.0&96.3\\
		&20&87.4 &94.7 &88.9 & 95.6& 89.2 & 96.6& 90.4& 97.0\\
		\hline
		
		\multirow{5}{*}{\textbf{Babble}}&
		0 & 62.4&80.2 & 63.8& 82.1& 65.7 &83.5 & 66.6 & 83.9\\
		&5& 76.2&87.3 & 77.6& 88.7& 79.4 & 89.1& 80.1 &90.3\\
		&10& 81.4&92.2 & 82.3& 93.5& 84.0 & 94.9&84.8 &94.5\\
		&15& 84.0&92.6 & 86.1& 94.0& 87.2 & 95.2& 89.1&95.7\\
		&20& 85.8&92.9 & 86.6& 95.1&  88.4 & 95.7 &90.3 &96.2\\
		\hline
		
		\textbf{Original}&  &88.5 &95.9 & 89.7&96.4 & 90.2 & 96.8& 91.1&97.7\\
		\hline
	\end{tabular}
	
	\caption{Comparison of speaker identification performances obtained using four different methods in various noise conditions.}
	\label{sid}
\end{table*}

\begin{table*}[h]
	\renewcommand{\multirowsetup}{\centering}  
	\renewcommand\arraystretch{1.0}
	\setlength{\tabcolsep}{2.8mm}
	\centering  
	\footnotesize
	\begin{tabular}{c|c|c|c|c|c|c|c|c|c}
		\hline
		\multirow{2}{*}{\textbf{Noise Type}}& \multirow{2}{*}{\textbf{SNR}} & \multicolumn{2}{c|}{\textbf{SID}}&\multicolumn{2}{c|}{\textbf{VoidID-Loss\cite{shon2019voiceid}}} &  
		\multicolumn{2}{c|}{\textbf{SESR-Step1}}&
		\multicolumn{2}{c}{\textbf{SESR-Step2}}\\
		
		\cline{3-10}
		& & EER (\%)&DCF&EER (\%)&DCF& EER (\%)& DCF& EER (\%)& DCF\\

		\hline
		\multirow{5}{*}{\textbf{Noise}}&
		0 &16.94 &0.993 &16.56 &0.938 & 16.02 &0.885 & 15.89 & 0.886\\
		&5& 12.48&0.855 &12.26 &0.830 & 11.87 & 0.794& 11.83 &0.786\\
		&10& 10.03&0.760 & 9.86&0.747 & 9.21& 0.695& 9.17 &0.695\\
		&15& 8.84&0.648 & 8.69&0.686 & 8.18 & 0.625 &7.99 &0.616\\
		&20& 7.96&0.594 & 7.83&0.639 & 7.06& 0.590 & 6.85&0.589\\
		\hline
		
		\multirow{5}{*}{\textbf{Music}}&
		0 & 17.04&0.940 & 16.24&0.913 & 15.69 & 0.893& 15.70 &0.904\\
		&5& 11.54&0.828 & 11.44&0.818 & 10.88 & 0.754& 10.78 &0.770\\
		&10& 9.69&0.749 & 9.13&0.733 & 8.76& 0.690& 8.94&0.704\\
		&15& 8.40&0.689 & 8.10&0.677 & 7.81& 0.631& 7.65&0.621\\
		&20& 7.70&0.665 & 7.48&0.635 & 7.09 & 0.606& 7.03&0.592\\
		\hline
		
		\multirow{5}{*}{\textbf{Babble}}&
		0 &38.90 &1.000 &37.96 &1.000 & 37.18 &0.999 & 37.52&0.994 \\
		&5& 28.04&0.998 & 27.12&0.996 & 26.84 & 0.991&26.69& 0.991\\
		&10& 17.34&0.917 & 16.66&0.926 & 16.38& 0.878 &16.93 &0.901\\
		&15& 11.31&0.795 & 11.25&0.807 & 10.87& 0.781 &10.84 &0.780\\
		&20&9.12 &0.720 & 8.99&0.705 & 8.76 & 0.679& 8.72&0.685\\
		\hline
		
		\textbf{Original}&  &6.92 &0.565 & 6.79&0.574 &6.52 & 0.548 & 6.48 &0.537\\
		\hline
	\end{tabular}
	
	\caption{Comparison of speaker verification performances obtained using four different methods in various noise conditions.}
	\label{sveri}
\end{table*}

\vspace*{-3mm}
\section{Results and Analysis}\label{results}
\vspace*{-2mm}

Table \ref{sid} shows the speaker identification performances obtained using two baselines (SID and VoiceID-Loss)
and two proposed approaches (SESR-Step1, SESR-Step2).
It is clear that the two proposed approaches, SESR-Step1 and SESR-Step2
can yield better performances than other baselines
in various noise conditions, even if the SNR is 0dB.
Moreover, after using speaker information learned by the SR1 module,
the proposed approach SESR-Step2
can further improve the identification performance in comparison with SESR-Step1.
This case is probably related to two factors.
The first factor is the use of speech enhancement before speaker identification and a joint optimization,
by which some noise interferences might be filtered out.
The second factor is the implementation of the speaker dependent speech enhancement in Step2.
Unlike speaker-independent speech enhancement, the use of speaker information
can not only recover the noise corrupted speech signals to some extent, but also possible highlights
speaker-specific features, which might be key to speaker recognition.

Table \ref{sveri} shows the speaker verification performances obtained using the four methods.
Similar to Table \ref{sid}, the use of SESR-Step2 can achieve the best results
in most conditions. 
When evaluating the verification performance, any further post-process, such as Probabilistic Linear Discriminant Analysis (PLDA) \cite{kenny2013plda},
was not employed, and only a cosine score was used to compare the similarities between enrolment and test data.
This might be the reason that the improvement of using SESR-Step2 over
SESR-Step1 on the speaker verification task is relative slight.

%
%

\begin{table}[tb]
	\renewcommand\arraystretch{1.0}
	\setlength{\tabcolsep}{0.5mm}
	\centering  
	\footnotesize
	\begin{tabular}{c|c|c|c|c}
		\hline
		SNR~~&~~Noisy~~&~~VoiceID-loss \cite{shon2019voiceid} ~~&~~ SESR-Step1 ~~&~~ SESR-Step2 \\
		\hline
		0 &1.53 &1.48 & 1.62  &1.90\\
		5& 1.78&1.72 & 1.89& 2.14\\
		10 & 1.86&1.83 & 1.97&2.35\\
		15& 2.16& 2.06 & 2.21 &2.58\\
		20& 2.39&  2.20& 2.53&2.89\\
		\hline
		
	\end{tabular}
	\caption{Comparison of the PESQ values obtained using the proposed approaches and baselines in various music noise conditions.}
	\label{PESQ}
\end{table}

\begin{table}[tb]
	\renewcommand\arraystretch{1.0}
	\setlength{\tabcolsep}{0.5mm}
	\centering  
	\footnotesize
	\begin{tabular}{c|c|c|c|c}
		\hline
		SNR~~&~~Noisy~~& ~~VoiceID-loss \cite{shon2019voiceid}~~& ~~SESR-Step1~~& ~~SESR-Step2 \\
		\hline
		0 & 0.53& 0.50 &0.56 & 0.63\\
		5 & 0.60& 0.58 & 0.64& 0.71\\
		10 & 0.65&  0.61& 0.67& 0.75\\
		15& 0.67&0.62& 0.69& 0.77\\
		20& 0.68& 0.64& 0.70& 0.78\\
		\hline
		
	\end{tabular}
	\caption{Comparison of the STOI values obtained using the proposed approaches and baselines in various music noise conditions.}
	\label{STOI}
\end{table}

Table \ref{PESQ} and \ref{STOI} show the speech enhancement performances
evaluated using PESQ and STOI respectively.
The second column in both tables show the quality of input speech corrupted by music
noise at five different SNR levels.
The third column indicates the obtained speech quality after using VoiceID-loss.
The use of proposed approach, SESR-Step2, shows clear advantages over 
VoiceID-loss and SESR-Step1 in various noise conditions. 

To further verify the robustness of the proposed approach against noise,
Figure \ref{fig:enhanced-spec} shows fours spectrograms: The noise corrupted speech by music noise at 0 dB; The enhanced speech obtained using SESR-Step1;
The enhanced speech obtained using SESR-Step2; The original speech.
It can be found that the music noise can be removed to a certain extent
from the spectrograms shown in Figure (b) and (c)
after using speech enhancement, and the spectrogram shown in figure (c) 
is closer to the original spectrogram shown in figure (d).  

\vspace*{-2mm}
\section{Conclusion and Future Work}\label{conclusion}
\vspace*{-1mm}
In this paper, a novel speaker-dependent speech enhancement for speaker recognition approach
is presented and tested on Voxceleb1. The obtained results show that
the use of the proposed approach can yield better performances in speaker recognition
and enhance speech quality in various noise conditions.

To further improve speaker identification performance and its robustness
against noise, more advanced deep learning technologies, such as capsule networks
and the vector quantizer variable auto-encoder (VQVAE) will tested in this framework.
In addition, some post-process methods, such as PLDA, will be also taken into account.
 

\vspace*{-2mm}
\begin{center}
	\large{\textbf{Acknowledgement}}
\end{center}
\vspace*{-2mm}
This work was in part supported by Innovate UK Grant number 104264.

\newpage
\bibliographystyle{IEEEtran}

\bibliography{mybib}

\begin{thebibliography}{10}
\providecommand{\url}[1]{#1}
\csname url@samestyle\endcsname
\providecommand{\newblock}{\relax}
\providecommand{\bibinfo}[2]{#2}
\providecommand{\BIBentrySTDinterwordspacing}{\spaceskip=0pt\relax}
\providecommand{\BIBentryALTinterwordstretchfactor}{4}
\providecommand{\BIBentryALTinterwordspacing}{\spaceskip=\fontdimen2\font plus
\BIBentryALTinterwordstretchfactor\fontdimen3\font minus
  \fontdimen4\font\relax}
\providecommand{\BIBforeignlanguage}[2]{{%
\expandafter\ifx\csname l@#1\endcsname\relax
\typeout{** WARNING: IEEEtran.bst: No hyphenation pattern has been}%
\typeout{** loaded for the language `#1'. Using the pattern for}%
\typeout{** the default language instead.}%
\else
\language=\csname l@#1\endcsname
\fi
#2}}
\providecommand{\BIBdecl}{\relax}
\BIBdecl

\bibitem{poddar2017speaker}
A.~Poddar, M.~Sahidullah, and G.~Saha, ``Speaker verification with short
  utterances: a review of challenges, trends and opportunities,'' \emph{IET
  Biometrics}, 2017.

\bibitem{variani2014deep}
E.~Variani, X.~Lei, E.~McDermott, I.~L. Moreno, and J.~Gonzalez-Dominguez,
  ``Deep neural networks for small footprint text-dependent speaker
  verification,'' in \emph{ICASSP}.\hskip 1em plus 0.5em minus 0.4em\relax
  IEEE, 2014.

\bibitem{snyder2018x}
D.~Snyder, D.~Garcia-Romero, G.~Sell, D.~Povey, and S.~Khudanpur, ``X-vectors:
  Robust dnn embeddings for speaker recognition,'' in \emph{ICASSP}.\hskip 1em
  plus 0.5em minus 0.4em\relax IEEE, 2018.

\bibitem{rahman2018attention}
F.~R. rahman Chowdhury, Q.~Wang, I.~L. Moreno, and L.~Wan, ``Attention-based
  models for text-dependent speaker verification,'' in \emph{ICASSP}.\hskip 1em
  plus 0.5em minus 0.4em\relax IEEE, 2018.

\bibitem{an2019deep}
N.~N. An, N.~Q. Thanh, and Y.~Liu, ``Deep cnns with self-attention for speaker
  identification,'' \emph{IEEE Access}, 2019.

\bibitem{leglaive2019speech}
S.~Leglaive, U.~{\c{S}}im{\c{s}}ekli, A.~Liutkus, L.~Girin, and R.~Horaud,
  ``Speech enhancement with variational autoencoders and alpha-stable
  distributions,'' in \emph{ICASSP 2019-2019 IEEE International Conference on
  Acoustics, Speech and Signal Processing (ICASSP)}.\hskip 1em plus 0.5em minus
  0.4em\relax IEEE, 2019, pp. 541--545.

\bibitem{sadeghi2019audio}
M.~Sadeghi, S.~Leglaive, X.~Alameda-Pineda, L.~Girin, and R.~Horaud,
  ``Audio-visual speech enhancement using conditional variational
  auto-encoder,'' \emph{arXiv preprint arXiv:1908.02590}, 2019.

\bibitem{zhao2014robust}
X.~Zhao, Y.~Wang, and D.~Wang, ``Robust speaker identification in noisy and
  reverberant conditions,'' \emph{IEEE/ACM Transactions on Audio, Speech, and
  Language Processing}, vol.~22, no.~4, pp. 836--845, 2014.

\bibitem{jang2017enhanced}
I.~Jang, C.~Ahn, J.~Seo, and Y.~Jang, ``Enhanced feature extraction for speech
  detection in media audio.'' in \emph{INTERSPEECH}, 2017, pp. 479--483.

\bibitem{farahani2006robust}
G.~Farahani, S.~M. Ahadi, and M.~M. Homayounpour, ``Robust feature extraction
  of speech via noise reduction in autocorrelation domain,'' in
  \emph{International Workshop on Multimedia Content Representation,
  Classification and Security}.\hskip 1em plus 0.5em minus 0.4em\relax
  Springer, 2006, pp. 466--473.

\bibitem{ming2007robust}
J.~Ming, T.~J. Hazen, J.~R. Glass, and D.~A. Reynolds, ``Robust speaker
  recognition in noisy conditions,'' \emph{IEEE Transactions on Audio, Speech,
  and Language Processing}, vol.~15, no.~5, pp. 1711--1723, 2007.

\bibitem{nahma2019adaptive}
L.~Nahma, P.~C. Yong, H.~H. Dam, and S.~Nordholm, ``An adaptive a priori snr
  estimator for perceptual speech enhancement,'' \emph{EURASIP Journal on
  Audio, Speech, and Music Processing}, vol. 2019, no.~1, p.~7, 2019.

\bibitem{yao2016priori}
R.~Yao, Z.~Zeng, and P.~Zhu, ``A priori snr estimation and noise estimation for
  speech enhancement,'' \emph{EURASIP journal on advances in signal
  processing}, vol. 2016, no.~1, p. 101, 2016.

\bibitem{ouyang2019fully}
Z.~Ouyang, H.~Yu, W.-P. Zhu, and B.~Champagne, ``A fully convolutional neural
  network for complex spectrogram processing in speech enhancement,'' in
  \emph{ICASSP 2019-2019 IEEE International Conference on Acoustics, Speech and
  Signal Processing (ICASSP)}.\hskip 1em plus 0.5em minus 0.4em\relax IEEE,
  2019, pp. 5756--5760.

\bibitem{pascual2017segan}
S.~Pascual, A.~Bonafonte, and J.~Serr{\`a}, ``Segan: Speech enhancement
  generative adversarial network,'' \emph{Proc. Interspeech 2017}, pp.
  3642--3646, 2017.

\bibitem{muhammed2019non}
P.~Muhammed~Shifas, N.~Adiga, V.~Tsiaras, and Y.~Stylianou, ``A non-causal
  fftnet architecture for speech enhancement,'' \emph{Proc. Interspeech 2019},
  pp. 1826--1830, 2019.

\bibitem{yao2019coarse}
J.~Yao and A.~Al-Dahle, ``Coarse-to-fine optimization for speech enhancement,''
  \emph{Proc. Interspeech 2019}, pp. 2743--2747, 2019.

\bibitem{shon2019voiceid}
S.~Shon, H.~Tang, and J.~Glass, ``Voiceid loss: Speech enhancement for speaker
  verification,'' \emph{Proc. Interspeech 2019}, pp. 2888--2892, 2019.

\bibitem{meng2019speaker}
Z.~Meng, Y.~Gaur, J.~Li, and Y.~Gong, ``Speaker adaptation for attention-based
  end-to-end speech recognition,'' \emph{arXiv preprint arXiv:1911.03762},
  2019.

\bibitem{zhao2018domain}
Y.~Zhao, J.~Li, S.~Zhang, L.~Chen, and Y.~Gong, ``Domain and speaker adaptation
  for cortana speech recognition,'' in \emph{2018 IEEE International Conference
  on Acoustics, Speech and Signal Processing (ICASSP)}.\hskip 1em plus 0.5em
  minus 0.4em\relax IEEE, 2018, pp. 5984--5988.

\bibitem{pandey2019new}
A.~Pandey and D.~Wang, ``A new framework for cnn-based speech enhancement in
  the time domain,'' \emph{IEEE/ACM Transactions on Audio, Speech, and Language
  Processing}, vol.~27, no.~7, pp. 1179--1188, 2019.

\bibitem{liu2018denoising}
J.-Y. Liu and Y.-H. Yang, ``Denoising auto-encoder with recurrent skip
  connections and residual regression for music source separation,'' in
  \emph{2018 17th IEEE International Conference on Machine Learning and
  Applications (ICMLA)}.\hskip 1em plus 0.5em minus 0.4em\relax IEEE, 2018, pp.
  773--778.

\bibitem{hajibabaei2018unified}
M.~Hajibabaei and D.~Dai, ``Unified hypersphere embedding for speaker
  recognition,'' \emph{arXiv preprint arXiv:1807.08312}, 2018.

\bibitem{willmott2005advantages}
C.~J. Willmott and K.~Matsuura, ``Advantages of the mean absolute error (mae)
  over the root mean square error (rmse) in assessing average model
  performance,'' \emph{Climate research}, vol.~30, no.~1, pp. 79--82, 2005.

\bibitem{nagrani2017voxceleb}
A.~Nagrani, J.~S. Chung, and A.~Zisserman, ``Voxceleb: a large-scale speaker
  identification dataset,'' \emph{arXiv preprint arXiv:1706.08612}, 2017.

\bibitem{snyder2015musan}
D.~Snyder, G.~Chen, and D.~Povey, ``Musan: A music, speech, and noise corpus,''
  \emph{arXiv preprint arXiv:1510.08484}, 2015.

\bibitem{cheng2004method}
J.-M. Cheng and H.-C. Wang, ``A method of estimating the equal error rate for
  automatic speaker verification,'' in \emph{2004 International Symposium on
  Chinese Spoken Language Processing}.\hskip 1em plus 0.5em minus 0.4em\relax
  IEEE, 2004, pp. 285--288.

\bibitem{van2007introduction}
D.~A. Van~Leeuwen and N.~Br{\"u}mmer, ``An introduction to
  application-independent evaluation of speaker recognition systems,'' in
  \emph{Speaker classification I}.\hskip 1em plus 0.5em minus 0.4em\relax
  Springer, 2007, pp. 330--353.

\bibitem{xie2019utterance}
W.~Xie, A.~Nagrani, J.~S. Chung, and A.~Zisserman, ``Utterance-level
  aggregation for speaker recognition in the wild,'' in \emph{ICASSP 2019-2019
  IEEE International Conference on Acoustics, Speech and Signal Processing
  (ICASSP)}.\hskip 1em plus 0.5em minus 0.4em\relax IEEE, 2019, pp. 5791--5795.

\bibitem{rix2001perceptual}
A.~W. Rix, J.~G. Beerends, M.~P. Hollier, and A.~P. Hekstra, ``Perceptual
  evaluation of speech quality (pesq)-a new method for speech quality
  assessment of telephone networks and codecs,'' in \emph{2001 IEEE
  International Conference on Acoustics, Speech, and Signal Processing.
  Proceedings (Cat. No. 01CH37221)}, vol.~2.\hskip 1em plus 0.5em minus
  0.4em\relax IEEE, 2001, pp. 749--752.

\bibitem{taal2010short}
C.~H. Taal, R.~C. Hendriks, R.~Heusdens, and J.~Jensen, ``A short-time
  objective intelligibility measure for time-frequency weighted noisy speech,''
  in \emph{2010 IEEE international conference on acoustics, speech and signal
  processing}.\hskip 1em plus 0.5em minus 0.4em\relax IEEE, 2010, pp.
  4214--4217.

\bibitem{he2016deep}
K.~He, X.~Zhang, S.~Ren, and J.~Sun, ``Deep residual learning for image
  recognition,'' in \emph{Proceedings of the IEEE conference on computer vision
  and pattern recognition}, 2016, pp. 770--778.

\bibitem{kingmaadam}
D.~P. Kingma and J.~L. Ba, ``Adam: Amethod for stochastic optimization.''

\bibitem{kenny2013plda}
P.~Kenny, T.~Stafylakis, P.~Ouellet, M.~J. Alam, and P.~Dumouchel, ``Plda for
  speaker verification with utterances of arbitrary duration,'' in \emph{2013
  IEEE International Conference on Acoustics, Speech and Signal
  Processing}.\hskip 1em plus 0.5em minus 0.4em\relax IEEE, 2013, pp.
  7649--7653.

\end{thebibliography}


\end{document}